# Ultrasensitive Textile Strain Sensors Redefine Wearable Silent Speech Interfaces with High Machine Learning Efficiency

**Authors**

Chenyu Tang[†,1], Muzi Xu[†,1], Wentian Yi[†,1], Zibo Zhang[2], Edoardo Occhipinti[3], Chaoqun Dong[1], Dafydd Ravenscroft[1], Sung-Min Jung[1], Sanghyo Lee[4], Shuo Gao[*,5], Jong Min Kim[1], Luigi G. Occhipinti[*,1]

**Affiliations**

[1]Electrical Engineering Division, Department of Engineering, University of Cambridge, UK

[2]Department of Electronic and Electrical Engineering, University College London, UK

[3]UKRI CDT in AI for Healthcare, Department of Computing, Imperial College London, UK

[4]School of Materials Science and Engineering, Kumoh National Institute of Technology (KIT), South Korea

[5]School of Instrumentation Science and Optoelectronic Engineering, Beihang University, China

[†]These authors contributed equally: Chenyu Tang, Muzi Xu, Wentian Yi
[*]Correspondence to: lgo23@cam.ac.uk, shuo_gao@buaa.edu.cn

**Abstract:** Designing silent speech interfaces (SSI) as wearable systems for real-world applications has been a challenging task for human-machine interface technologies due to the need to simultaneously fulfil three key requirements: device comfort and durability, high time-energy efficiency in signal decoding, and high accuracy in speech decoding. Our research introduces an SSI technology that meets all these criteria in real-world application scenarios. We developed a biocompatible strain sensor integrated into a comfortable textile choker that molds to the throat's contours, and is capable of stably enduring over 10,000 stretching-releasing cycles. The proposed sensing mechanism is founded on the reversible formation of ordered microcracks of graphene flakes embedded in textile substrates, resulting in unprecedented levels of sensitivity. The measured gauge factor exceeds existing state-of-the-art strain sensors by 420%, which allow us to adeptly capture the most subtle movements of the throat during speaking, breathing or other tasks involving the throat cavity and muscles. This enhanced sensitivity simplifies signal processing demands in sharp contrast with traditional approaches that rely on complex voice recognition algorithms. The resulting signals are characterized by high information density, eliminating the need for heavy multidimensional data analyses, and enabling decoding of speech signals using a computationally efficient neural network model. Our network's backbone is based on a one-dimensional convolutional neural network with residual structures, forgoing traditional filters and instead employing real-world noise injections to samples to enhance model robustness. This end-to-end compact design leads to considerable time and energy savings, with the network reducing the computational load by 90% while maintaining a high accuracy of 95.25% for a lexicon of 20 words from 3 participants. Crucially, the network swiftly adapts to new users and words, demonstrating robust generalization with as few as 30 samples per class and maintaining a decoding accuracy of 90%. Our research demonstrates the practicality of a wearable SSI that blends high sensitivity and precision with ease of use, advancing the field toward real-time communication systems for daily applications.

# I. Introduction

Silent speech interfaces (SSI) have emerged as a cutting-edge solution for scenarios where verbal communication is hindered. These include environments with excessive noise that can significantly interfere with spoken language, or cases involving physiological conditions such as stroke, cerebral palsy, Parkinson's disease, or recovery from laryngeal surgeries [1, 2]. By analyzing nonvocal human signals, SSI offers a novel method for decoding speech in silent conditions. Among the various challenges in SSI research, developing an effective wearable system for real-world applications is a key objective for researchers. To achieve this goal, it is crucial to ensure that the device is comfortable and durable enough for practical use to encourage user acceptance. Additionally, it is vital that the system operates with high precision and efficiency in distinguishing the speech of different users across a variety of scenarios.

In recent years, researchers have been actively working to develop effective SSI systems suitable for real-world wearable applications. This involves the innovation of new devices for capturing human silent speech signals and the design of improved algorithmic models. Human speech-related neural impulses originate in the central nervous system, travel through the peripheral nervous system to the vocal cords and are then articulated with the help of facial movements, resulting in various speech sounds [3]. In pursuit of decoding this complex process, scientists have developed a range of SSI systems. For instance, techniques such as electroencephalography (EEG) [4, 5, 6] and electrocorticography (ECoG) [7, 8, 9] have been employed to decode speech from brain activity. However, these methods, while innovative, often fall short in practicality for wearable technology due to their invasive nature and the complexity of their setups.

Simultaneously, in the quest to create more user-friendly SSI, researchers have turned to analyzing mechanical movements in the throat and face, employing sensors such as electromyography (EMG) [10, 11, 12] and strain sensors [13, 14, 15, 16, 17]. These approaches show promise for integration into wearable devices, being noninvasive and more adaptable to daily use. Although this shift toward physical signal detection has theoretically enhanced wearability, it still faces its own set of challenges, notably the delicate balance between user comfort, signal accuracy, and system efficiency (Supplementary Figure 1). Enhanced user comfort often necessitates fewer sensory channels, reducing the wearable burden on the human body, which can lead to less detailed data capture and reduced accuracy in speech decoding. To mitigate this, an increase in the complexity of the data processing models is needed, such as increasing the system's sampling rate to capture more nuanced speech nuances or converting signals into two-dimensional images to enhance data richness, but this solution raises issues of computational load, affecting the overall system efficiency [18, 19]. This interdependence between the three aspects - comfort, accuracy, and efficiency - creates a significant hurdle in the development of practical, wearable SSI systems.

Bridging this gap requires innovative solutions that ensure user comfort without compromising the accuracy and efficiency of the system, a challenge that lies at the heart of current SSI research for real-world wearable applications.

In our research, we address the challenges of wearable SSI with a unique blend of advanced sensor design and computational efficiency. We have developed an ultrasensitive textile sensor, that is intricately integrated into a wearable choker. This sensor, characterized by its high information density signals and complemented by a matching lightweight end-to-end neural network, expertly balances user comfort with high precision and system efficiency (Figure 1a). The distinctive sensing mechanism is based upon its unique structure, featuring ordered microcracks on graphene-coated textiles, which significantly enhances sensitivity (Figure 1b). In silent speech scenarios, particularly within small strain ranges, our sensor achieves a gauge factor improvement of 420% over previous best efforts (Figure 1c). This exceptional sensitivity enables the capture of information-rich speech signals, allowing for their efficient processing through our specially designed lightweight neural network, which achieves a high accuracy of 95.25% while concurrently reducing the network's computational load by 90%. This approach negates the need for the high-dimensional, complex model augmentations often associated with traditional SSI algorithms. Our custom one-dimensional convolutional neural network processes this dense information effectively, thus reducing the computational load while maintaining high accuracy. This groundbreaking synergy of sensor design and neural network optimization not only bridges the gap between user convenience and technical efficacy but also sets a new standard in wearable silent speech communication technologies, forging avenues for seamless, natural communication across diverse settings.

## II. Results

**Textile Strain Sensor based on Ordered Cracks**

To capture abundant information for eliminating the need for laborious multidimensional analyses, high sensitivity within small sensing strain ranges ($\leqslant 5\%$) [20, 21] is an indispensable property of flexible wearable sensors developed for detecting the throat micromovements associated with speech. Based on the fact that speaking different words corresponds to different degrees of stretching or shrinking strains of the throat muscle [22, 23], features for words hidden in the signals can be extracted by enhancing the sensitivity of the strain sensors. Our proposed ultrasensitive textile strain sensor with ordered cracks possesses the ability to detect tiny deformations of throat skin and distinguish the fundamental signal characteristics even among words with extremely similar pronunciations. Due to exceptionally high sensitivity resulting from ordered cracks formed on the surface of the textile substrate, high-density information can be provided for word recognition.

With numerous unique features, including conformability, breathability and durability, the textile is considered an ideal substrate for human motion monitoring with extraordinary performance [24, 25]. However, the resistance change of traditional textile strain sensors fabricated by printing/coating methods with relatively low gauge factor within a small strain range is insufficient to capture adequate information required for decoding different words, as shown in the inset of Figure 1c. Therefore, we developed a structured graphene sensing layer with ordered cracks, which dramatically improves the sensitivity of the textile strain sensor (Figure 2a). Such ordered cracks can be formed through one-step printing. By increasing the number of printing layers of graphene ink, graphene is not only printed on the surface of a single fiber but also forms a continuous layer of graphene on the top of the textile substrate. Due to the stiffness mismatch between the top graphene layer and the textile substrate, a series of ordered cracks can be created by utilizing the textile matrix as the template after prestretching (Figure 2b and Figure 2c). When no strain is applied on the sensor, these ordered crack edges return to contact. As the strain increases gradually, the distance between these ordered cracks becomes larger, and the contact areas decrease rapidly, leading to a sharp change in contact resistance. Then, the textile strain sensor is equipped with the capability to sense the tiny deformation generated by throat micromovements because the large change in contact areas introduced by ordered cracks magnifies the resistance change with a small strain applied, and the gauge factor can reach 317 within 5% strain. Moreover, ordered cracks are beneficial to the stability of the resistance response in comparison with the reduction in stability due to the nonuniformity of cracks that exist randomly in the graphene layer with a certain thickness [26].

In addition to the ultrahigh sensitivity brought by the ordered cracks, the fabrication method of our textile strain sensors is biocompatible, simple, low cost, and scalable, and the property and performance can be easily controlled by tuning the parameters of the manufacturing process. Owing to its defects that are advantageous for piezoresistivity [27], graphene nanoplatelets are used in the preparation of functional ink (DI-water based) through high-pressure homogenization, a straightforward method that weakens the van der Waals forces between graphite layers [28]. Figure 2d shows the aspect ratio distribution of graphene flakes we used with a mean value of ~ 45. By altering the size of the interaction chamber of the homogenizer, the aspect ratio of nanoplatelets, which influences the percolation threshold [29], can be adjusted (Supplementary Figure 3). Screen printing is renowned for its customizable pattern, exceptional compatibility with a flexible substrate, affordable cost and scalable fabrication in the field of printing electronics [30, 31, 32]. Diverse patterns on the printing mesh can be transferred on our textile substrate (made from 95% bamboo fibers and 5% elastane) directly, and varying the number of printing layers can control the thickness of the ordered cracks.

The performance of our textile strain sensor with ordered cracks was evaluated by monitoring the variations in its relative resistance. Within a small sensing range, the textile strain sensor demonstrates a linear relative resistance response with relatively low hysteresis (Figure 2e). Figure 2f displays the stable stretching-releasing responses under 1%, 1.5%, 2%, 3%, 4% and 5% strain, and the relative resistance increases linearly with strain ($\leqslant 5\%$), showing the high reliability of the sensor within a small strain range. Meanwhile, this textile strain sensor exhibits the ability to resist tensile frequency interference (Figure 2g), which would be useful for identifying the same word spoken with different pitches. The detection limit was tested, as shown in Figure 2h. Based on ordered cracks, the textile strain sensor realizes an ultralow detection limit (0.05%), which is crucial for tiny strain detection. The durability determines the lifespan of the sensor applied in the real world, and our textile strain sensor can endure over 10000 stretching-releasing cycles with stable and reliable electrical functionality (Figure 2i). Such excellent durability is mainly ascribed to the outstanding adhesion between graphene ink and substrates through the selection of additives for ink and preprocessing of textiles by plasma treatment and the remarkable stability of ordered cracks formed along the textile matrix under repeated stretching-releasing. These overall characteristics of the textile strain sensor with ordered cracks pave the way for real-world application of the silent speech system.

**The Lightweight End-to-end Neural Network for Robust Speech Recognition**

In general, various SSI systems based on EMG sensors or strain sensors mainly encounter three types of noise in real-world applications: flicker noise caused by sensor imperfections, sound noise from the external environment, and physiological noise arising from users' bodily movements, such as breathing, swallowing, or neck movements, when wearing the device. Figure 3a shows a typical signal pattern during speech recognition using our smart choker. Initially, when the user is not wearing the choker, the signal collected by the readout module appears as a superposition of the DC offset, corresponding to the sensor's initial resistance and flicker noise. It is worth noting that at the fifth second, we introduced 100 dB environmental sound noise. From the response and our subsequent multiple tests on sound noise, it can be concluded that although our smart choker is extremely sensitive to the micromovement of the skin at the throat, it is 100% unresponsive to environmental sound noise. After the choker is worn, the DC offset changes, which is determined by the varying tightness with which the user wears the choker. After wearing, the noise in the signal appears as a superposition of flicker noise and physiological noise. Instead of using filters, we implemented noise injection data augmentation to enhance the system's noise immunity. Although previous methods such as additive Gaussian noise injections have significantly improved model robustness, we devised a simple "random noise window" technique to better assist the model in learning real-world noise characteristics (Figure 3c) [48]. Initially, users wear the choker silently, engaging in normal activities such as breathing and turning their heads. The signals collected during this time by the readout module represent a noise background without speech. We then randomly select multiple noise windows of the same length as speech samples and overlay the noise from these windows onto the speech samples to create augmented speech samples. This approach, compared to traditional filtering methods, greatly enhances energy efficiency. Such efficiency is vital for wearable systems in real-world applications, as it facilitates extended wearability without compromising performance.

Considering the high information density from our device's exceptional sensitivity, we crafted an end-to-end lightweight one-dimensional neural network for processing and classifying SSI signals. As shown in Figure 3d, our model unites a series of convolutional layers with fully connected layers, and each component is finely tuned to the subtleties of the SSI data. At the heart of our network are residual blocks, featuring pairs of one-dimensional convolutional layers with a kernel size of 3. This design ensures critical temporal feature capture while optimizing computational efficiency. Each convolutional layer incorporates batch normalization and ReLU activation to bolster stability and learning efficacy. The initial convolution layer, equipped with 64 size-7 filters, followed by batch normalization and ReLU activation, plays a pivotal role in initial feature extraction from input signals. A dropout layer with a 0.2 rate is integrated to mitigate overfitting and maintain robustness across diverse scenarios. Efficient data downsampling is achieved via max-pooling, aligning with our model's focus on handling consistent 3-second, 1500-point signal samples at 500 Hz, which is critical for precise, real-time SSI applications. Concluding the network architecture are the fully connected layers, leading to a classification layer adept at distinguishing specific speech words, reflecting the tailored design of our system for SSI-based communication. A detailed network structure can be found in Supplementary Figure 9.

In Figure 3e, our model demonstrates remarkable capability, achieving high accuracy in classifying the top 20 frequently used English words with outstanding time and energy efficiency compared to state-of-the-art systems, as characterized by low inference floating-point operations per second (FLOPS). This efficiency highlights our network's ability to harness single-channel, high-density data from our sensitive SSI device while minimizing computational demand. Such a streamlined approach promises extended wearability and practicality for daily use, establishing a new benchmark for energy-efficient silent speech recognition.

**Performance in Real-world Silent Speech Scenarios**

To validate the efficacy of our SSI system in real-world application scenarios, we collected three datasets (based on English) from three participants (see relevant details in Supplementary Table 2) across three of the most common speech communication settings. In Dataset 1, we gathered the ten most frequently used verbs and ten nouns in spoken English, using this collection as a baseline experiment to verify the system's capability to recognize words commonly used in everyday life [50]. For Dataset 2, we compiled a set of ten easily confusable word pairs that differ by only one phonetic

element—vowels, consonants, or stress—such as "book" and "look", "sheep" and "ship", and the verb and noun pronunciations of "record". In Dataset 3, we collected five lengthy words at varying reading speeds to test the system's ability to correctly decode the same word across different speech rates. The details of the vocabulary for the three datasets can be found in Supplementary Table 3, and Figure 4d provides a visualization of the signals for the word "Cambridge" at three different reading speeds.

In each of the three datasets, we collected 100 samples for every example, with 80 designated for the training set and 20 for the testing set. In Dataset 1, our model achieved a classification accuracy of 95.25% for the 20 high-frequency words (see the corresponding confusion matrix in Figure 4a); in Dataset 2, we reached a classification accuracy of 93% for the 10 confusable words (see the corresponding confusion matrix in Figure 4b); and in Dataset 3, our model achieved a classification accuracy of 96% for the five long words read at different speeds (see the corresponding confusion matrix in Figure 4d). To highlight the strengths of our network structure, we conducted a model evaluation on Dataset 1 (the baseline dataset), comparing our network with state-of-the-art benchmark backbones (all in 1D mode, results shown in Supplementary Figure 10). Our network demonstrated advantages in both accuracy and time and energy efficiency. Additionally, to investigate whether our lightweight network's simpler architecture could limit performance on larger datasets with more samples per class, we compared the accuracy achieved by models trained with varying numbers of samples (see Supplementary Figure 11). The results indicated that model accuracy continued to increase with more training samples, without reaching a saturation point, suggesting that the model's performance could be further optimized with the introduction of more data.

To assess whether our model exhibits bias in classification—such as focusing on noise or other irrelevant signal regions—we employed Relevance-Class Activation Mapping (R-CAM) to visualize the signal areas that the model concentrates on during classification (Figure 4c) [51]. The visualization reveals that the model consistently directs its attention to the key micromovements associated with the words, indicating a targeted and effective recognition process. Moreover, as demonstrated by several word examples in the figure, the DC offsets of the samples vary. This variation arises from our data collection strategy, which embraced the diversity of choker tightness and accounted for slight differences in placement with each wear. This diversity underscores the robustness of our system to the subtle variations in wear positioning and tightness that different users may exhibit in real-world scenarios, ensuring reliability across repeated uses.

To evaluate our system's performance on new users and words it has not encountered before, we utilized our baseline model trained on Dataset 1 as a pretrained model and transferred it to a new user and ten new words (Figure 5a). For the new user, we collected the same five words previously gathered from the original three participants. For the new words, we selected the ten confusable words from Dataset 2 as novel entries for the baseline model. We observed that our model could effectively recognize the new user and words with minimal fine-tuning: with only 15 to 20 samples per class, the model achieved an 80% accuracy rate for both new words and users, which is a 43% and 53% improvement, respectively, compared to training directly on new data without a pretrained model. With fine-tuning on just 30 samples per class, the model reached a decent 90% accuracy for both new users and words (Figure 5b). Figures 5c and 5d visualize the model's generalization performance on new users and words using t-SNE, showing a significant improvement in the model's classification capabilities after leveraging the learning experience of the baseline pretrained model. Notably, in Figure 5d, the model's ability to discriminate between confusable words, such as "book" and "look", is markedly enhanced, indicating the model's robust feature extraction capabilities and strong generalization.

## III. Conclusions

In this study, we introduce a groundbreaking ultrasensitive textile sensor, integrated into a wearable choker, revolutionizing the field of silent speech interfaces (SSI) for real-world applications. This sensor, enhanced with ordered microcracks on graphene-coated textiles, offers exceptional sensitivity and durability, significantly simplifying speech signal decoding. Coupled with a tailored, energy-efficient neural network, it demonstrates high accuracy and reduced computational load and is ideal for wearable technology in real-world scenarios. This innovative system excels in decoding a wide range of words, adapts swiftly to new users and vocabularies, and shows robustness against various noises and physical wear variations. This advancement in SSI technology paves the way for future applications in seamless, natural communication and broader human-computer interaction fields.

## IV. Methods

**Materials**

TIMREX KS 25 Graphite (synthetic graphite with a particle size of 25μm) was purchased from IMERYS. The sodium deoxycholate (SDC) (≥97%) and sodium carboxymethyl cellulose (CMC-Na: an average molecular weight of 700000), as the surfactant and the binder for ink preparation, were both obtained from SIGMA-ALDRICH. Deionized water was provided by PURELAB Flex Pure Water System. The textile substrate (95% bamboo fibers and 5% elastane) was purchased from Jelly Fabrics Ltd.

**Preparation of Graphene Ink**

The functional graphene ink for the sensor fabrication is prepared by high pressure homogenization (HPH), a liquid phase exfoliation (LPE) method to fabricate graphene, and the steps are illustrated as follows. First, as the surfactant, SDC is dissolved in deionized (DI) water (5g/L) to prevent aggregation of fillers by electrostatic repulsion. Second, TIMREX KS 25 graphite flakes were added to the SDC solution (100g/L) and mixed by dissolver at 500rpm for 30min. Then, the mixtures are exfoliated by a high-pressure homogenizer (PSI-40) using a dual slot deagglomeration chamber (D200D: 200μm). It is processed at a pressure of 700 bar and 70 exfoliation cycles. Finally, CMC-Na as a binder was added to the graphene dispersion (10g/L) to stabilize the flakes and control the viscosity of the printing ink, and the prepared ink was stirred for 3h at room temperature to fully dissolve CMC-Na.

**Fabrication of a Textile Strain Sensor with Ordered Cracks**

The textile graphene-based strain sensor with ordered cracks is composed of a functional graphene sensing film and a textile substrate providing flexibility and support. With this simple structure, the manufacturing process is straightforward via screen printing. First, the textile is treated by UV-ozone treatment (UV ozone cleaner UVC-1014, NanoBioAnalytics) for 30 min at room temperature to improve the hydrophilicity of the substrate and the adhesion between the graphene ink and textile. Then, with the cooperation of the screens (mesh count 90T: 230 mesh/inch) and squeegee, the 100g/L prepared graphene ink is squeezed on the substrate fixed on the platform of the printer with a rectangular pattern. This printing process is repeated 7 times to form a continuous graphene layer on the top of the textile substrate. After printing, ordered cracks are formed along the textile matrix by prestretching the sensor at 5% strain.

**Characterization of the structure and performance of the sensor**

The lateral size, thickness and aspect ratio of graphene flakes were assessed by Bruker Icon AFM (Supplementary Figure 3). One hundred flakes were measured from 3 AFM scans, each with scan area ~20 μm × 20 μm. SEM images were obtained using a Magellan 400 to characterize the morphology of the textile strain sensor with ordered cracks. Supplementary Figure 4 shows the SEM results of the fabrication process. A tensile testing machine (Deben Microtest 200N Tensile Stage, INSTRON universal testing system) and digital sourcemeter (Keithley 2400 Source Meter Unit) were used to measure the electromechanical properties of the textile strain sensor with ordered cracks. The resistance responses upon repetitive and consecutive strains were recorded to evaluate the sensing performance of the sensor.

**Experimental setup of data acquisition**

Our strain sensor is printed onto a choker, with copper tape tightly affixed to both ends of the sensor at a 1-centimeter distance, and a potentiostat (EmStat4S, PalmSens) is utilized as the readout module for data acquisition. The readout module inputs a voltage of 1V and outputs the current passing through the strain sensor. We selected a sampling frequency of 500Hz for the signal, with each word sample lasting 3 seconds. Supplementary Movie 1 offers a demonstration of the data collection process.

**Software environment**

The processing of the data and the training of the network were conducted in an environment based on Python 3.8.13, Miniconda 3, and PyTorch 2.0.1, with training acceleration provided by Apple's Metal Performance Shaders (MPS). During the noise injection phase, each original sample was augmented with real-world noise from four different random noise windows, creating four new samples. The optimal hyperparameters for model training can be found in Supplementary Table 4.

**Data availability**

The data supporting this study are available in https://doi.org/10.17863/CAM.104307

**Acknowledgments**

E.O. was supported by UKRI Centre for Doctoral Training in AI for Healthcare (grant No. EP/S023283/1), D.R. was supported by EPSRC Centre for Doctoral Training in Sensors Technologies and Applications (grant No. EP/L015889/1), S.G. acknowledges funding from National Natural Science Foundation of China (grant No. 62171014), S.L. acknowledges funding from National Research Foundation of Korea Grant funded by the Korean Government (NRF-2018R1A6A1A03025761), W.Y. was supported by Pragmatic Semiconductor (grant No. G117793), C.T. was supported by Endoenergy Systems (grant No. G119004), L.G.O. acknowledges funding from EPSRC (grants No. EP/K03099X/1, EP/L016087/1, EP/W024284/1, EP/P027628/1), the EU Graphene Flagship Core 3 (grant No. 881603), and Haleon (grant No. G110480). We would like to extend our sincere gratitude to Prof. George Malliaras for his invaluable guidance and mentorship throughout this work as the PhD advisor to C.T. and M.X.

# Figures

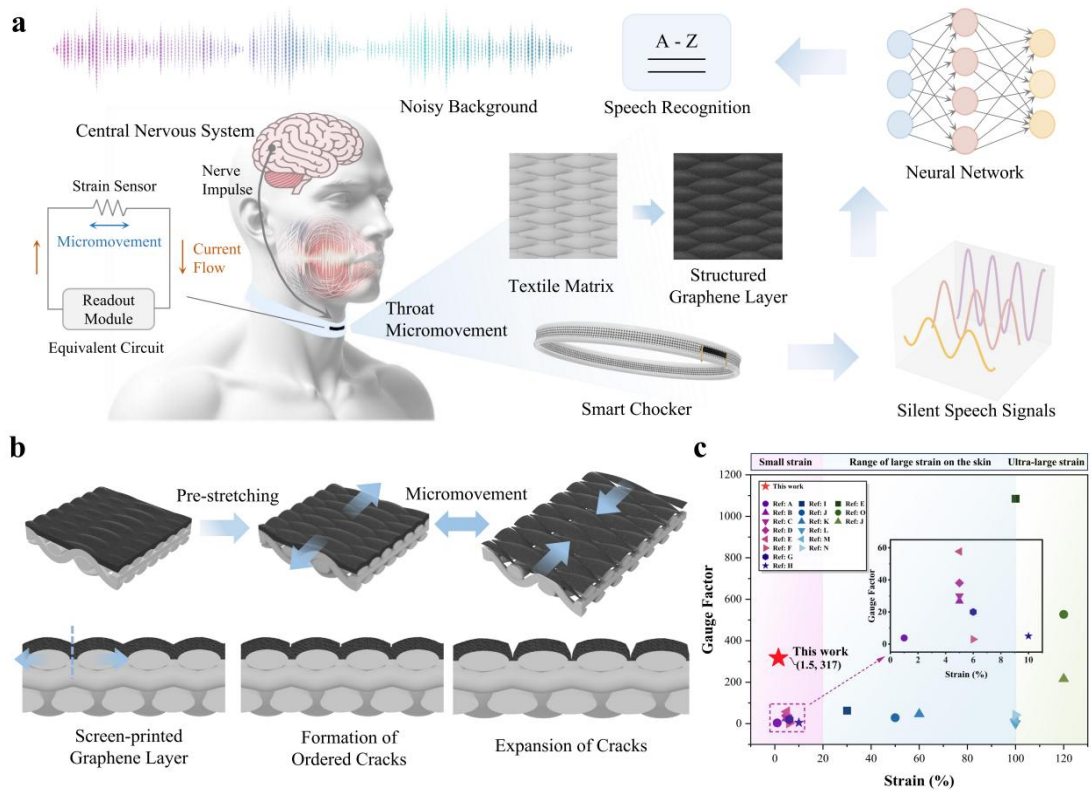

**Figure 1 | Comprehensive overview of the wearable SSI, featuring an ultrasensitive strain sensor and a neural network for efficient speech recognition. a,** The process of speech recognition initiates with nerve impulses from the central nervous system translating into micromovements in the throat. These movements are then captured by an ultrasensitive strain sensor integrated into a smart choker, comprising a textile substrate with an overlying structured graphene layer. The sensor responds by altering its resistance, resulting in a change in electrical signal, which is then captured and processed by a readout module. The obtained electrical signals are fed into a lightweight end-to-end neural network for processing and speech recognition. The detection of throat micromovement based on orderly-cracked graphene ensures robust performance even in noisy environments, leveraging the high resistance of the sensor to background interference. **b,** The sensing mechanism in the textile-based strain sensor enhanced with a structured graphene layer. This layer is created through the screen-printing of a continuous thick graphene film onto a textile matrix. Following a prestretching process, the inherent ordered weaving structure of the textile induces the formation of ordered thorough cracks in the graphene layer, which are strategically aligned with the weave. The structured graphene layer can dynamically respond to throat micromovements with significant and abrupt changes in electrical resistance. **c,** Comparative analysis showcasing the performance metrics of our printed textile-based graphene strain sensor against other reported strain sensors, focusing on the strain scale and Gauge factor. The exceptional gauge factor of our sensor in the small strain range is critical for capturing rich, information-dense signals. A-N refer to [33-47].

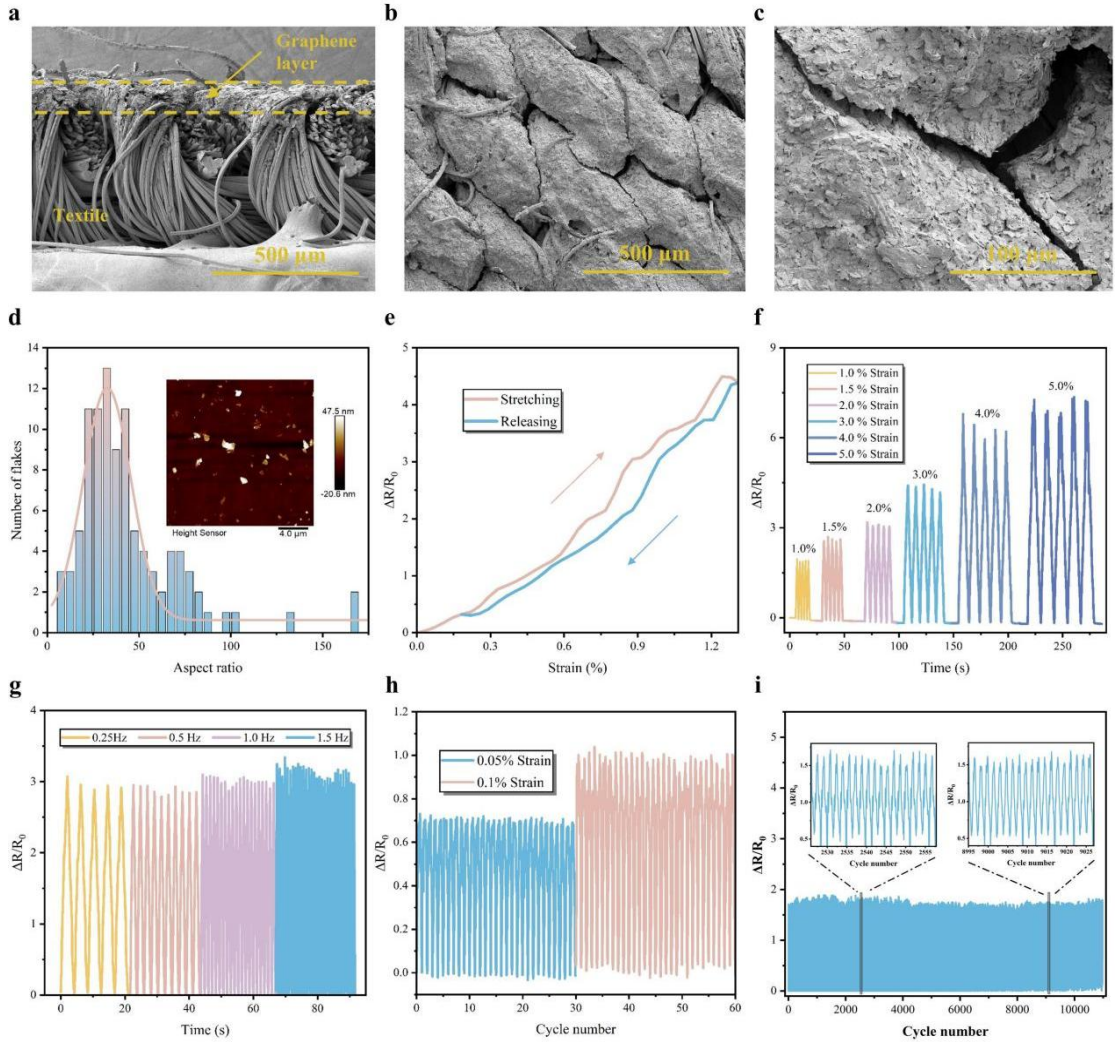

**Figure 2 | Characterization of the device. a,** Cross-sectional SEM image of the textile strain sensor with ordered cracks, showing the top graphene layer and the bottom textile layer. **b,** Top view SEM image of the ordered cracks formed on the top of the textile substrate. **c,** SEM image of the surface through-crack structure. **d,** Aspect ratio distribution of graphene flakes fabricated by the high pressure homogenizer. The inset shows an AFM image of graphene flakes. **e,** Hysteresis of the relative resistance change during a stretching-releasing cycle. **f,** Relative resistance responses with 1.0%, 1.5%, 2.0%, 3.0%, 4.0% and 5.0% cyclic strains. **g,** Relative resistance responses with different stretching-releasing rates under 1.5% cyclic strain. **h,** Detection limit stability test of the textile strain sensor under 0.05% and 0.1% cyclic strains. **i,** Durability test of the textile strain sensor with ordered cracks by multicyclic stretching and releasing with a strain of 1.5% over 10000 cycles.

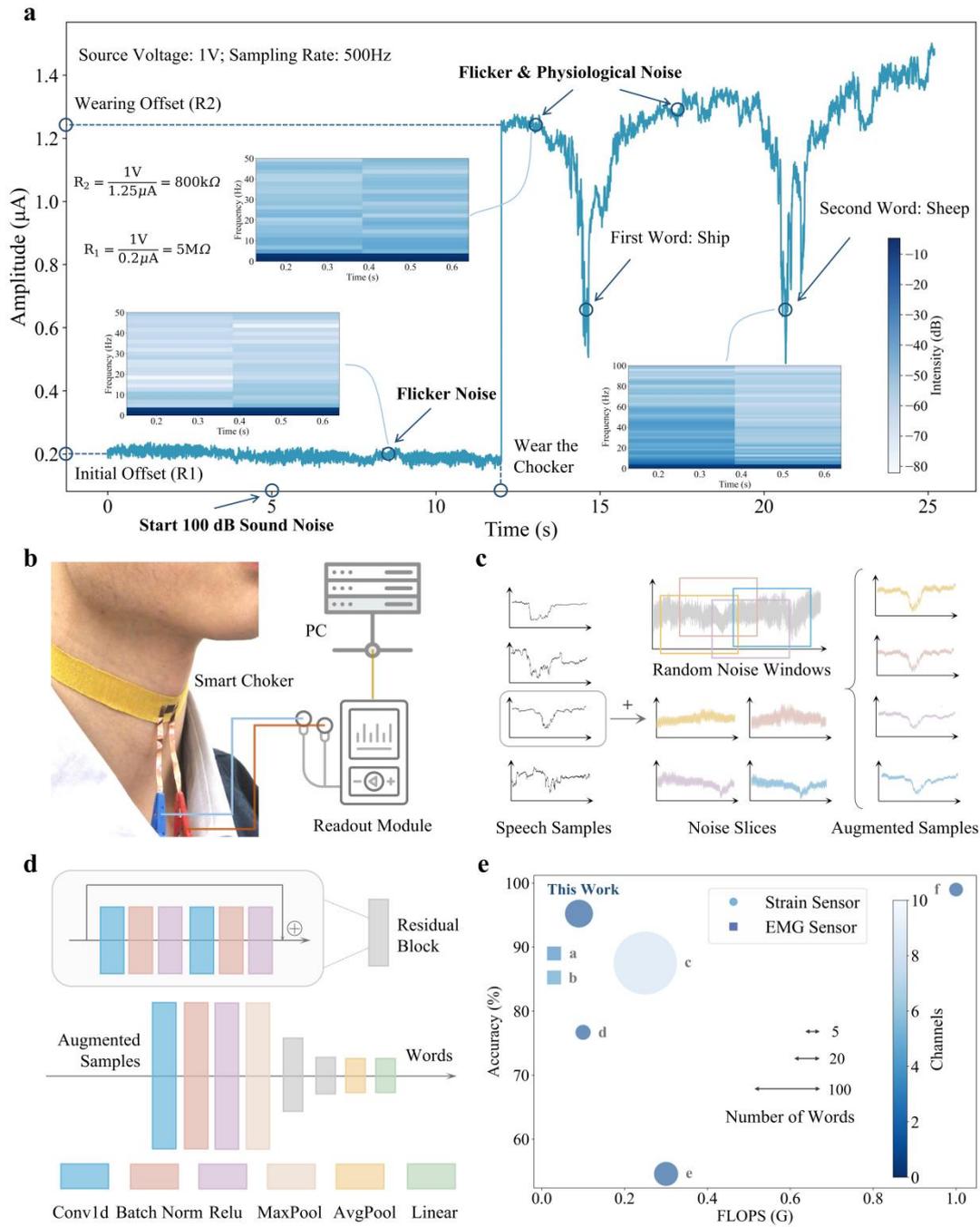

**Figure 3 | System and model architecture. a,** The signal characteristics of an entire silent speech phase are presented. At the 5th second, a sound noise of 100dB is introduced. Starting at the 12th second, a choker is worn, and signals of two words are collected. Three segments are extracted to visualize the spectrogram, illustrating intensity variations across different frequencies over time. **b,** Flowchart of the entire system, comprising the smart choker, readout module, and the PC for model processing. **c,** Flowchart depicting the random noise injection method used for data augmentation. **d,** Pipeline of the lightweight end-to-end neural network employing one-dimensional convolutional layers. **e,** Comparison of model efficiency (measured in FLOPS), accuracy, and channel usage with relevant works; a-f refer to [12, 13, 18, 49, 14, 19].

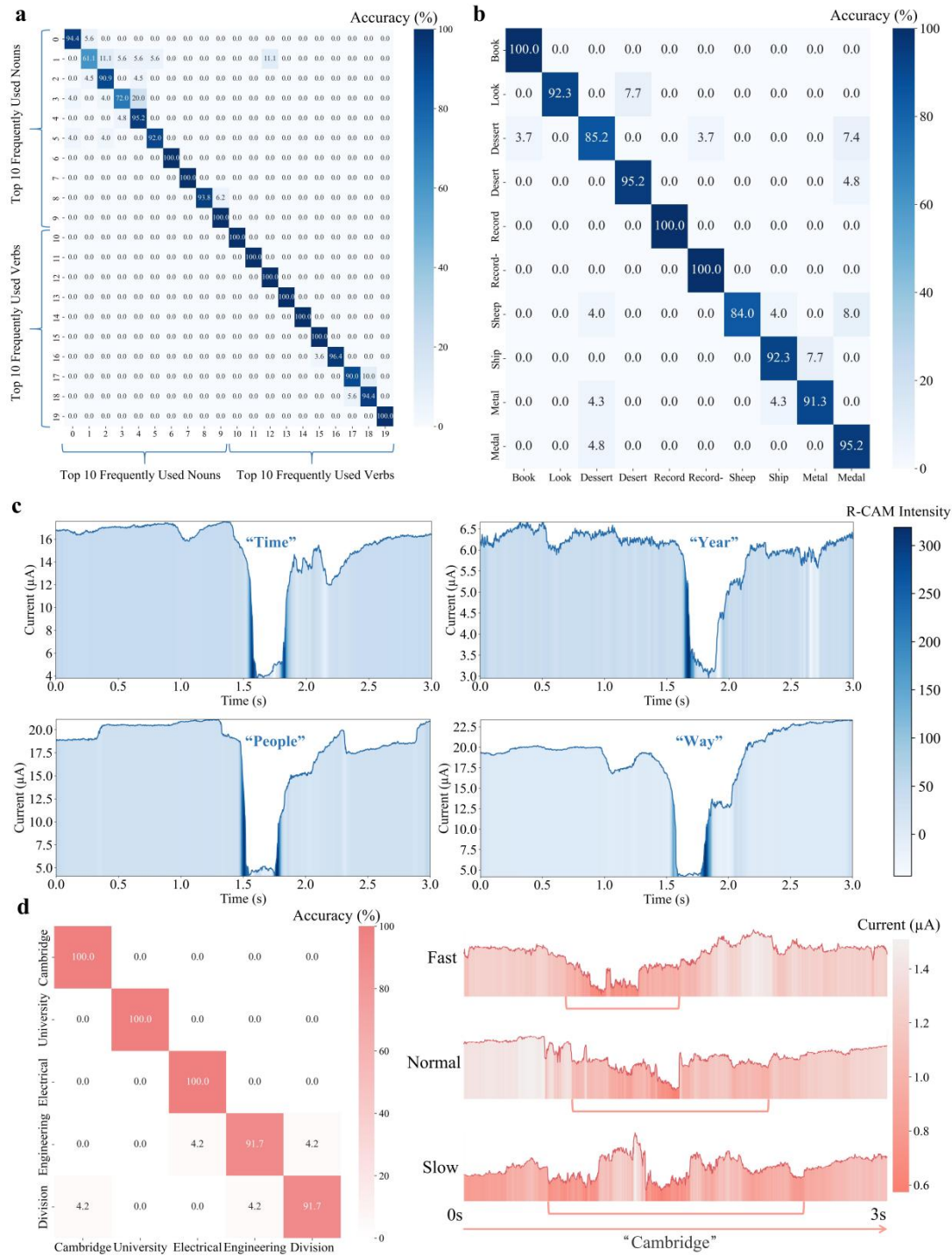

**Figure 4 | Silent speech recognition results. a,** Confusion matrix showing the classification results for the 10 most frequently used verbs and 10 most frequently used nouns, indicating the model's capability in everyday use. **b,** Confusion matrix for the classification of 10 words that are easily confused in terms of vowels, consonants, or stress patterns, demonstrating the model's ability to discern subtle differences. **c,** Relevance-Class Activation Mapping (R-CAM) utilized to highlight the signal areas the model focuses on during word classification. **d,** Confusion matrix for the classification of 5 long words read at varying speeds, showcasing the model's robustness to different reading speeds. **e,** Visualization of the long word "Cambridge" read at three different speeds.

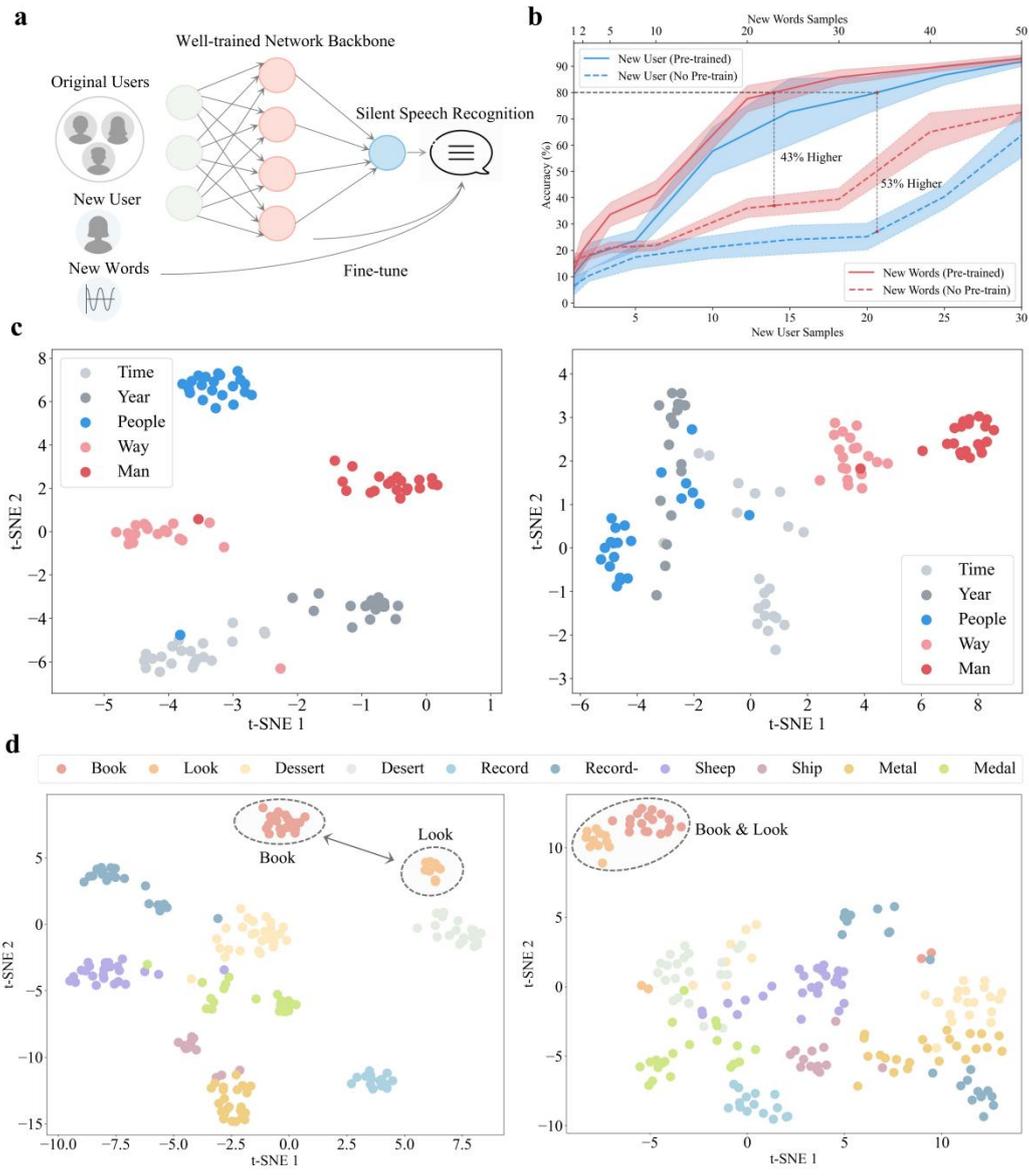

**Figure 5 | Generalization ability. a,** Flowchart depicting the model's generalization process. **b,** Evaluation results of the model's generalization capabilities: comparison of accuracies when trained from scratch and fine-tuned using a baseline model with samples from new users and new words in varying quantities. **c,** T-distributed stochastic neighbor embedding (t-SNE) visualizations comparing models trained from scratch with new user data (right) to those fine-tuned using a baseline model (left). **d,** T-SNE visualizations showing the difference in models trained from scratch with new word data (right) compared to those fine-tuned using a baseline model (left).

# Supplementary Information

**Supplementary Figure 1 | Comparison between our method and traditional methods.** Previous silent speech methods often faced a trade-off where it was difficult to simultaneously satisfy good user-friendliness, high recognition accuracy, and high system efficiency. However, our method, by using a single-channel ultrasensitive throat strain sensor, collects one-dimensional signals with high information density. The signals are learned by our specially designed lightweight, end-to-end one-dimensional neural network, realizing a wearable silent speech system framework that simultaneously possesses high user comfort, high speech recognition accuracy, and high system efficiency.

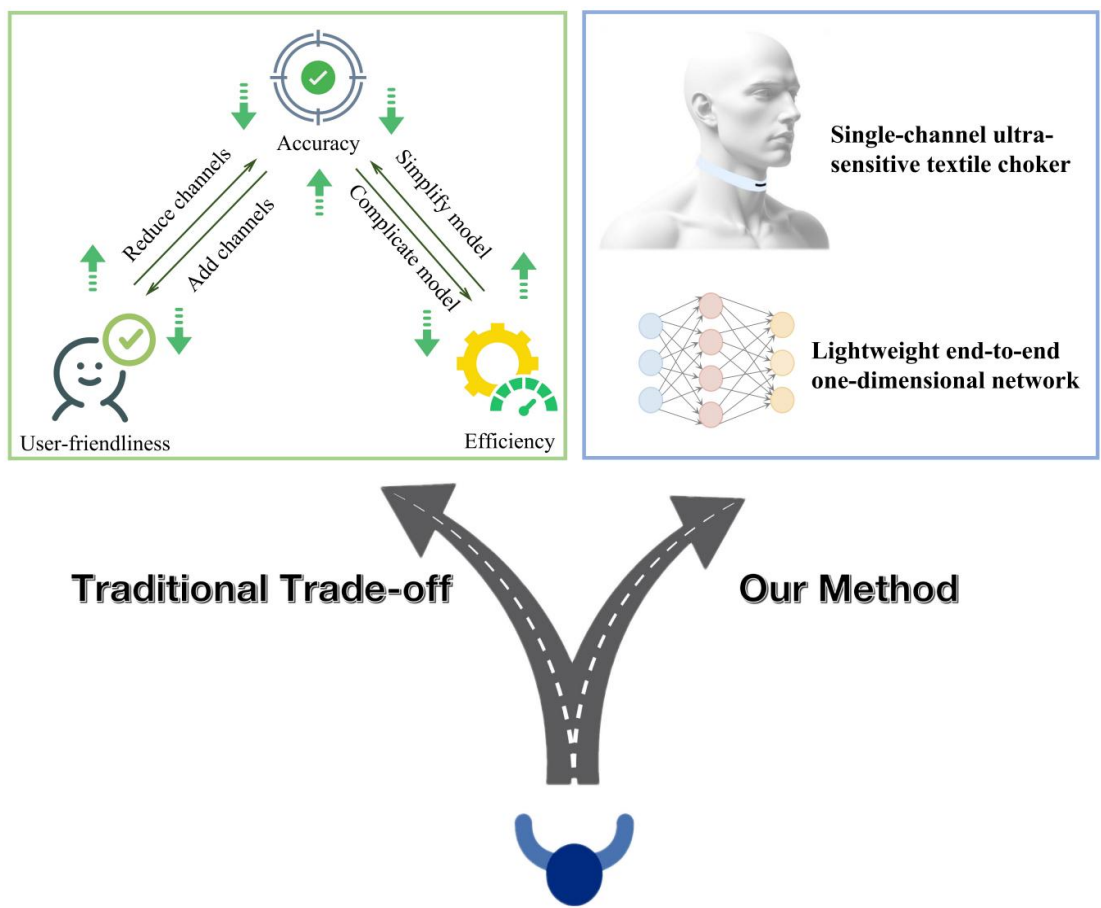

**Supplementary Figure 2 | Fabrication of the textile strain sensor with ordered cracks. a,** Preparation of graphene ink via high-pressure homogenization. The mixture of graphite and SDC passes through the microchannel of the interaction chamber under high pressure, where the graphite is exfoliated into graphene flakes at a high shear rate. Subsequently, CMC is added to the graphene dispersion to form the final graphene ink. **b,** Fabrication of a textile-based strain sensor via screen printing. The textile substrate is cleaned with UV ozone to enhance hydrophilicity. Then graphene ink is applied onto it with a stencil and squeegee. Typically, over 7 cycles of screen printing are needed to create a thick, continuous graphene layer. The device fabrication is completed by prestretching the sensor to generate ordered, thorough cracks.

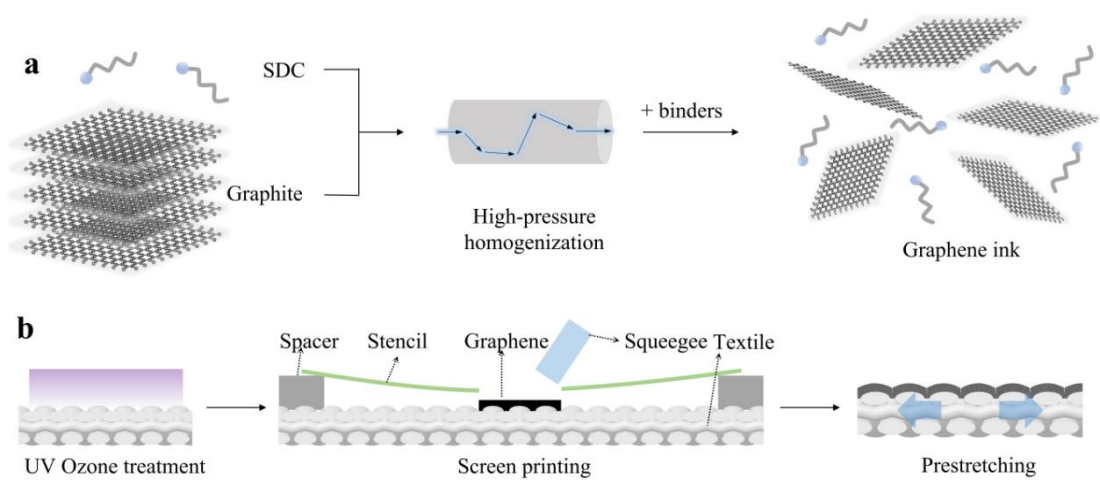

**Supplementary Figure 3 | Characterization of graphene flakes fabricated by high-pressure homogenizer with 200 μm and 87 μm interaction chambers. a-f:** Results of the graphene flakes prepared via 200 μm interaction chamber. **a-c,** AFM images (20 μm × 20 μm) of the graphene flakes at three different locations. **d-f,** Lateral size, thickness and aspect ratio distributions of graphene flakes based on 100 flakes randomly selected from three AFM scans. **g-l:** Results of the graphene flakes prepared via an 87 μm interaction chamber. **g-i,** AFM images (20 μm × 20 μm) of the graphene flakes at three different locations. **j-l,** Lateral size, thickness and aspect ratio distributions of graphene flakes based on 100 flakes randomly selected from three AFM scans. The graphene flakes prepared via 87 μm (with a mean value of 48) have a larger aspect ratio than 200 μm (with a mean value of 45).

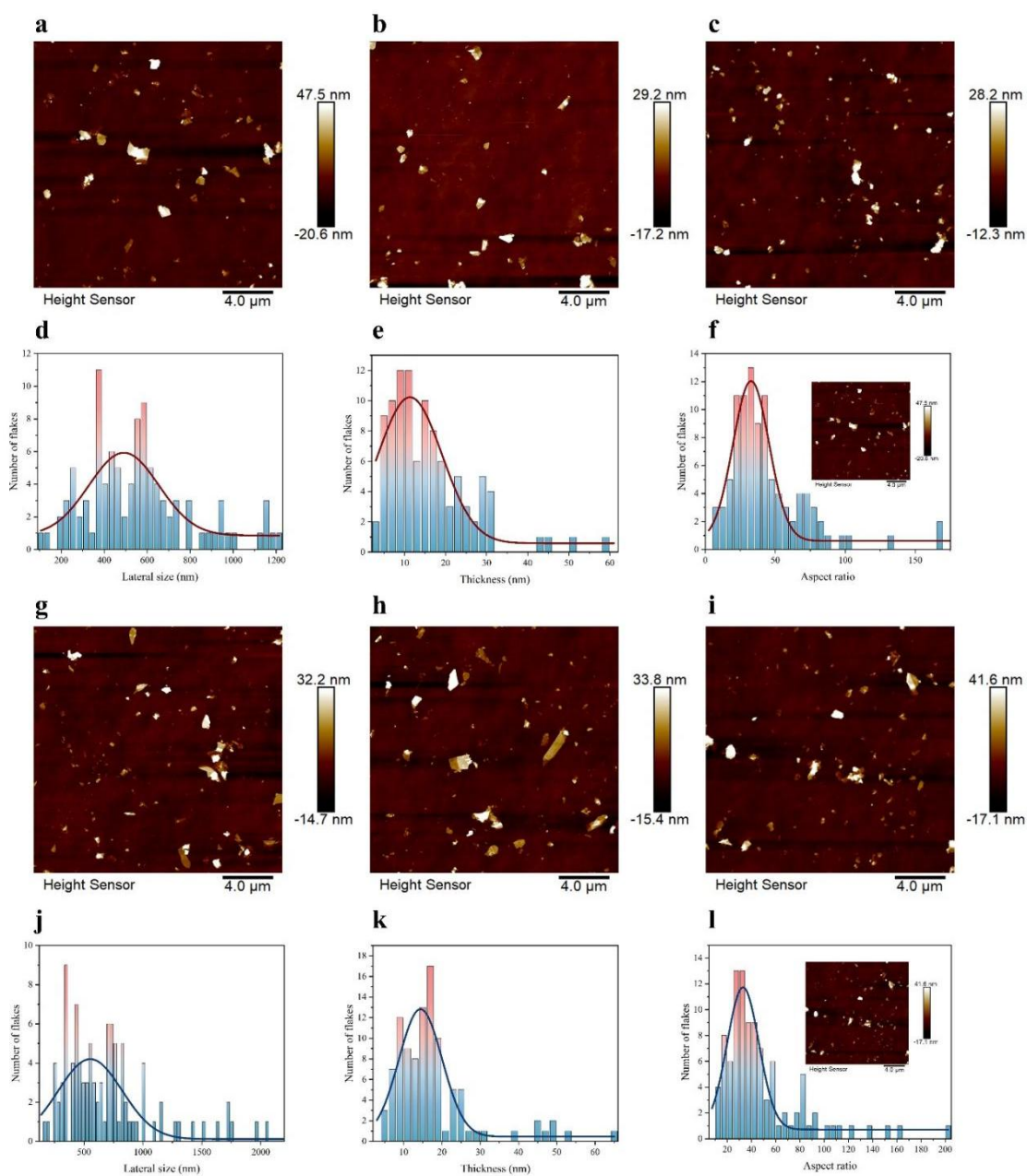

**Supplementary Figure 4 | SEM results of the fabrication process. a-b,** Top-view SEM images of the textile substrate with one and three printing layers. As the number of printing layers increases, a continuous layer of graphene will be formed on the top of the textile substrate as shown in Figure 2a,b (7 printing layers). **c-d,** SEM images of the single fiber without and with 7 printing layers of graphene. **e-f,** Cross-sectional and top-view SEM images of the ordered cracks.

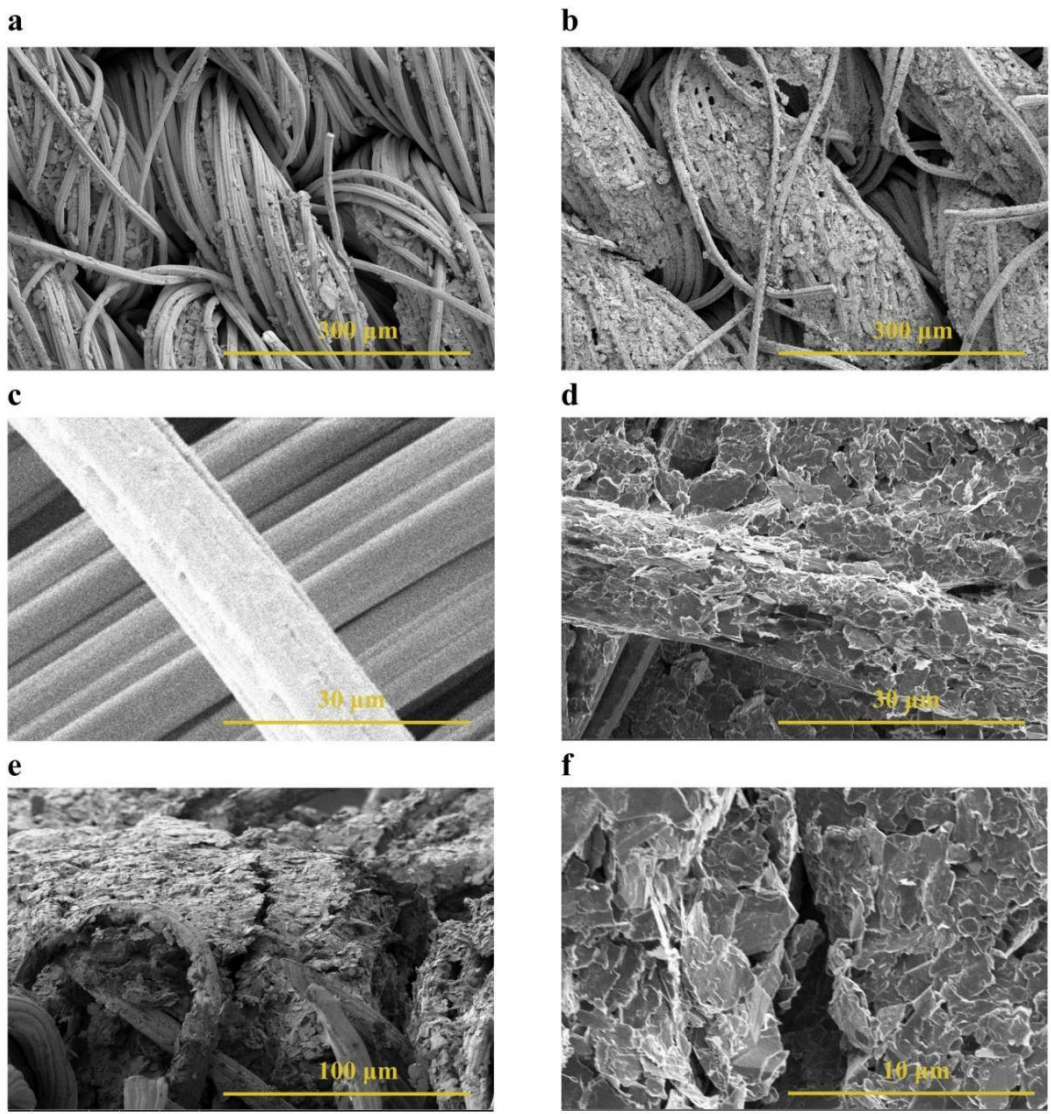

**Supplementary Fig. 5 | Original resistance signals of the 10 most frequently used nouns collected by the textile strain sensor with ordered cracks. a,** time. **b,** year. **c,** people. **d,** way. **e,** man. **f,** day. **g,** thing. **h,** child. **i,** Mr. **j,** government.

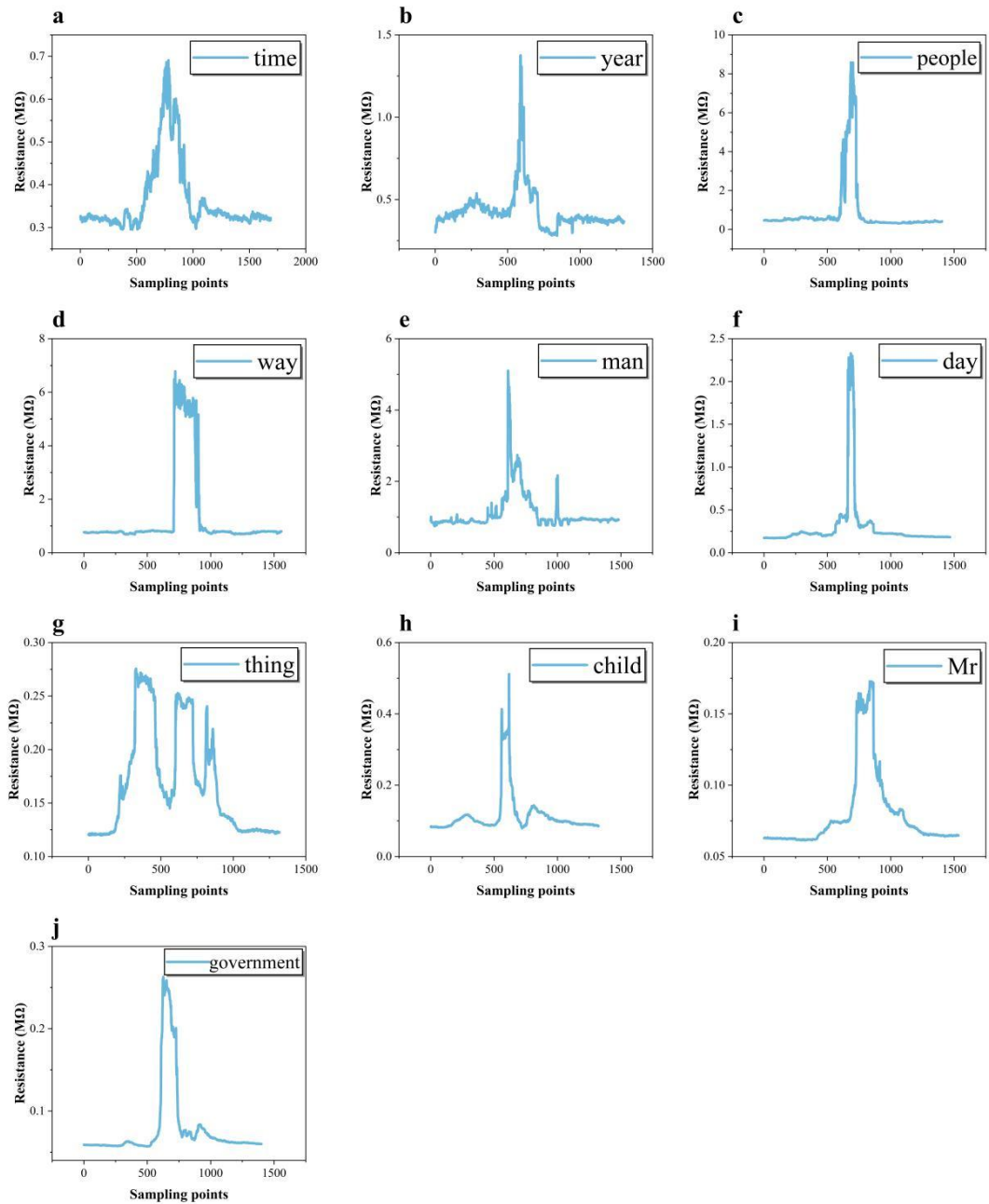

**Supplementary Fig. 6 | Original resistance signals of the 10 most frequently used verbs collected by the textile strain sensor with ordered cracks. a,** be. **b,** have. **c,** do. **d,** will. **e,** say. **f,** would. **g,** can. **h,** get. **i,** make. **j,** go.

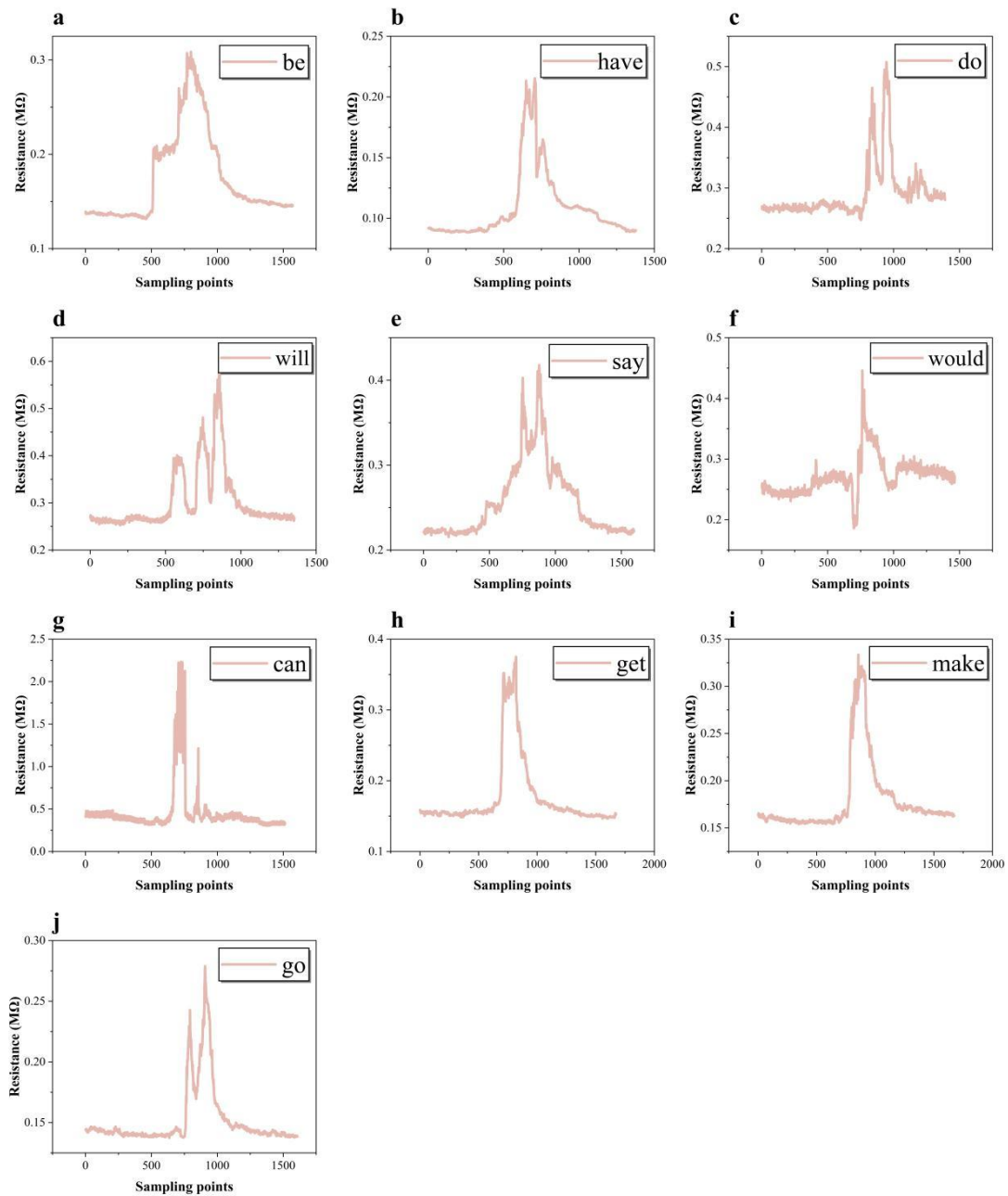

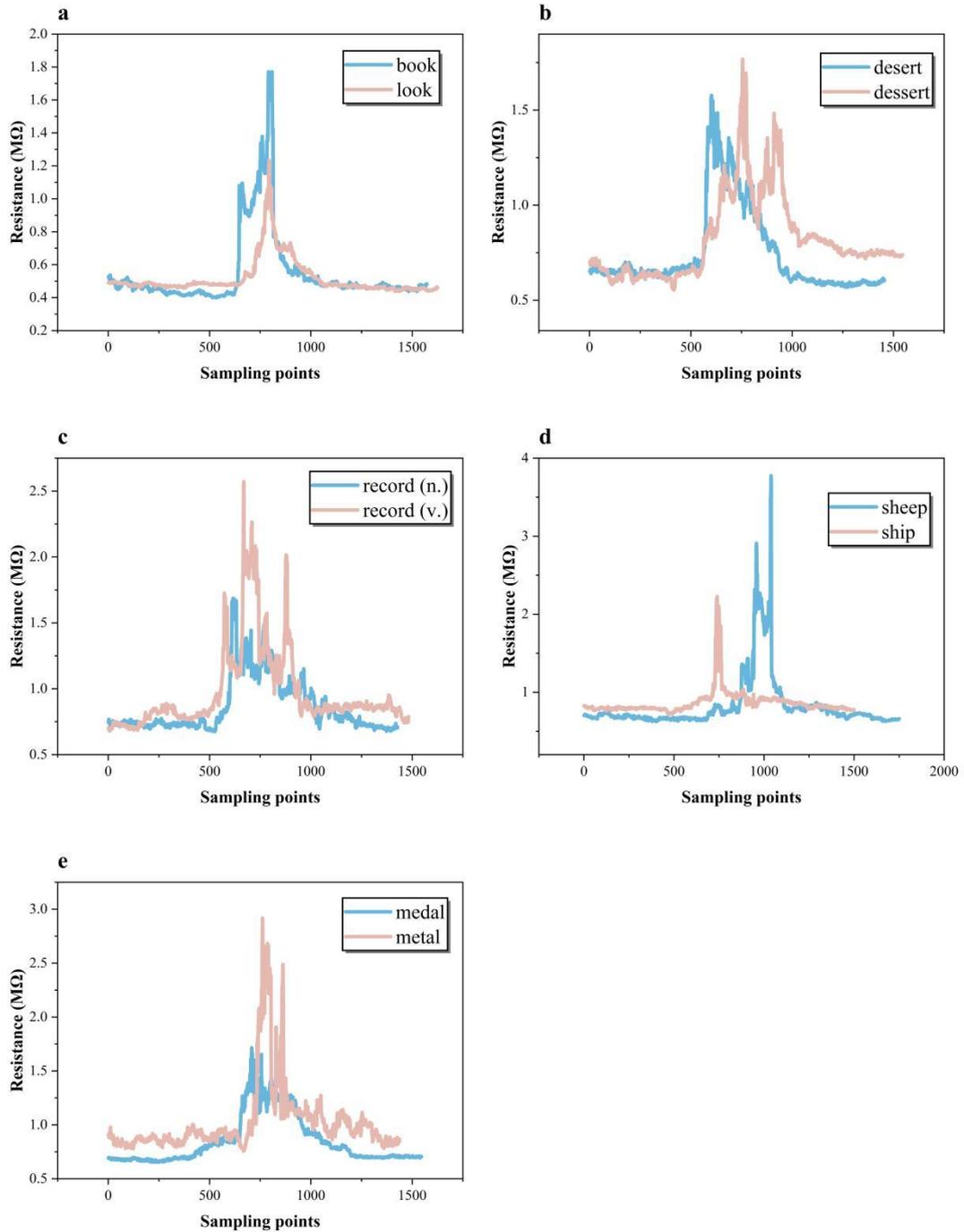

**Supplementary Fig. 7 | Original resistance signals of 5 pairs of words that are easily confused in terms of vowels, consonants, or stress patterns collected by the textile strain sensor with ordered cracks. a,** book & look. **b,** desert & dessert. **c,** record (n.) & record (v.). **d,** sheep & ship. **e,** medal & metal.

**Supplementary Fig. 8 | Original resistance signals of 5 long words read at varying speeds (fast, normal, slow). a,** Division. **b,** Cambridge. **c,** Engineering. **d,** Electrical. **e,** University.

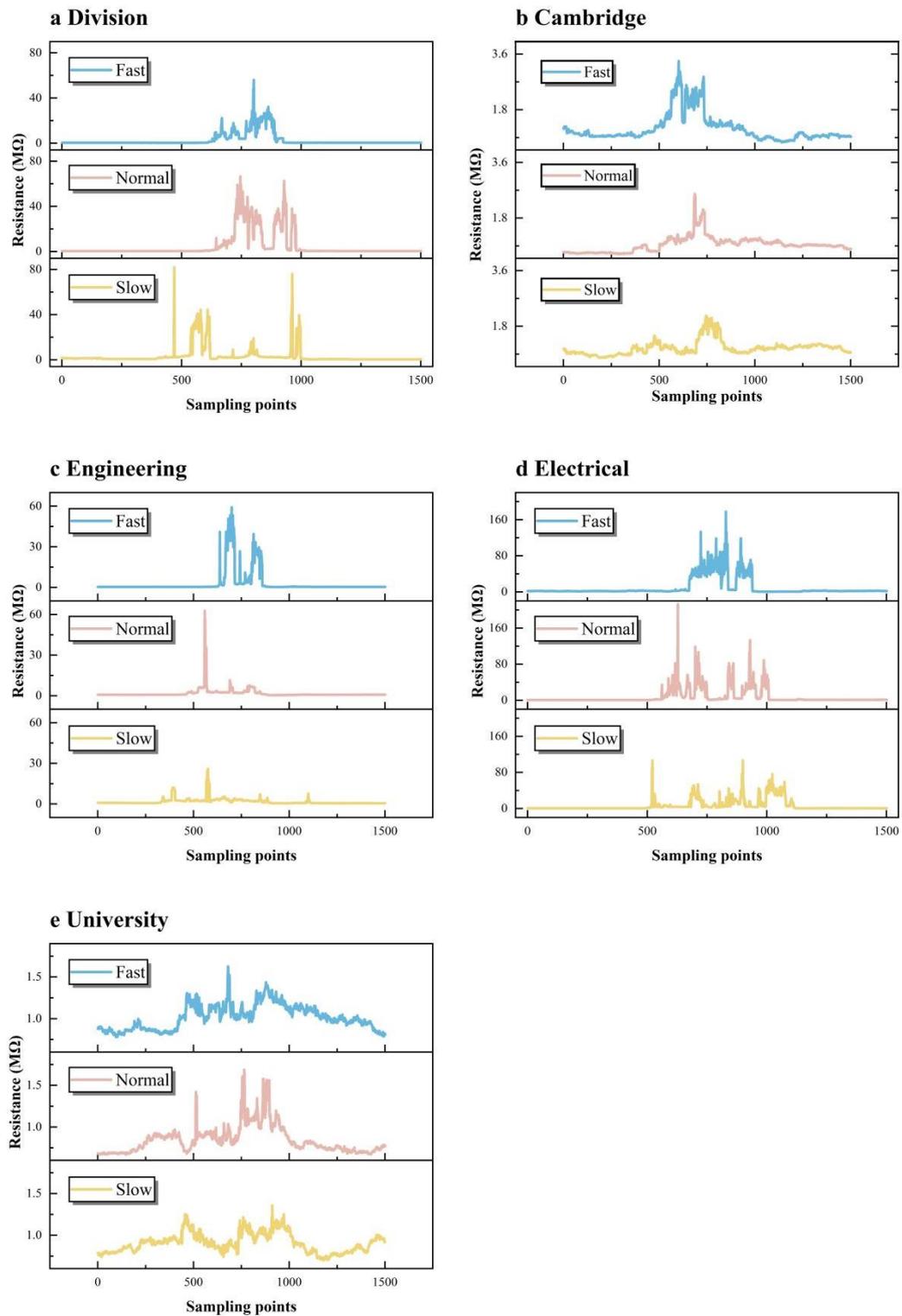

**Supplementary Figure 9 | Details of the model backbone.**

```
----------------------------------------------------------------
        Layer (type)               Output Shape         Param #
================================================================
            Conv1d-1              [-1, 64, 750]             512
       BatchNorm1d-2              [-1, 64, 750]             128
              ReLU-3              [-1, 64, 750]               0
         MaxPool1d-4              [-1, 64, 375]               0
            Conv1d-5             [-1, 128, 375]          24,704
       BatchNorm1d-6             [-1, 128, 375]             256
              ReLU-7             [-1, 128, 375]               0
            Conv1d-8             [-1, 128, 375]          49,280
       BatchNorm1d-9             [-1, 128, 375]             256
           Conv1d-10             [-1, 128, 375]           8,320
      BatchNorm1d-11             [-1, 128, 375]             256
             ReLU-12             [-1, 128, 375]               0
    ResidualBlock-13             [-1, 128, 375]               0
           Conv1d-14             [-1, 256, 188]          98,560
      BatchNorm1d-15             [-1, 256, 188]             512
             ReLU-16             [-1, 256, 188]               0
           Conv1d-17             [-1, 256, 188]         196,864
      BatchNorm1d-18             [-1, 256, 188]             512
           Conv1d-19             [-1, 256, 188]          33,024
      BatchNorm1d-20             [-1, 256, 188]             512
             ReLU-21             [-1, 256, 188]               0
    ResidualBlock-22             [-1, 256, 188]               0
AdaptiveAvgPool1d-23               [-1, 256, 1]               0
           Linear-24                    [-1, 20]           5,140
================================================================
Total params: 418,836
Trainable params: 418,836
Non-trainable params: 0
----------------------------------------------------------------
```

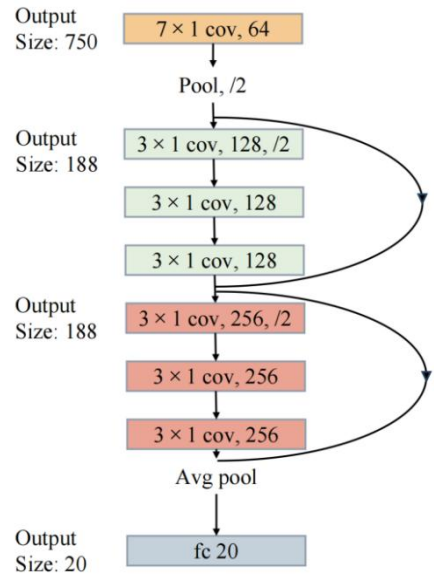

**Supplementary Figure 10 | Performance comparison to state-of-the-art benchmark neural networks (all in 1D version) on the baseline dataset.**

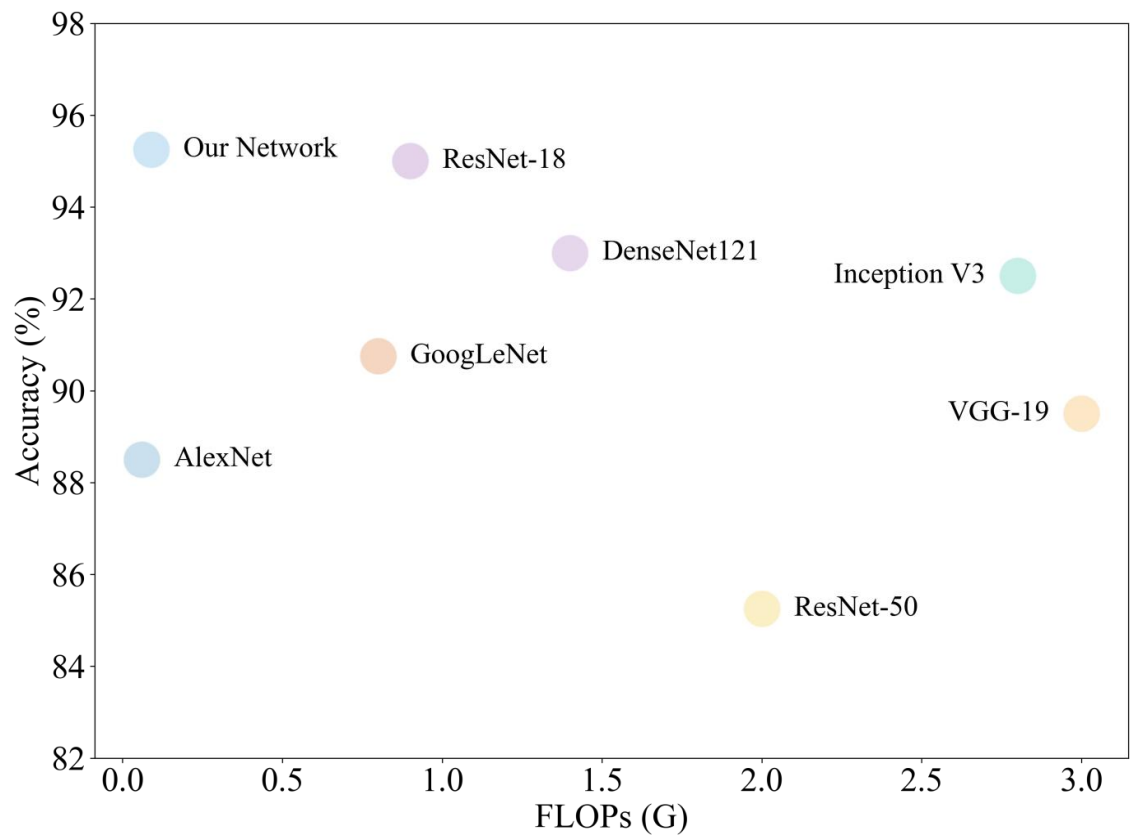

**Supplementary Figure 11 | Recognition performance of the baseline model after training with different numbers of samples per class.**

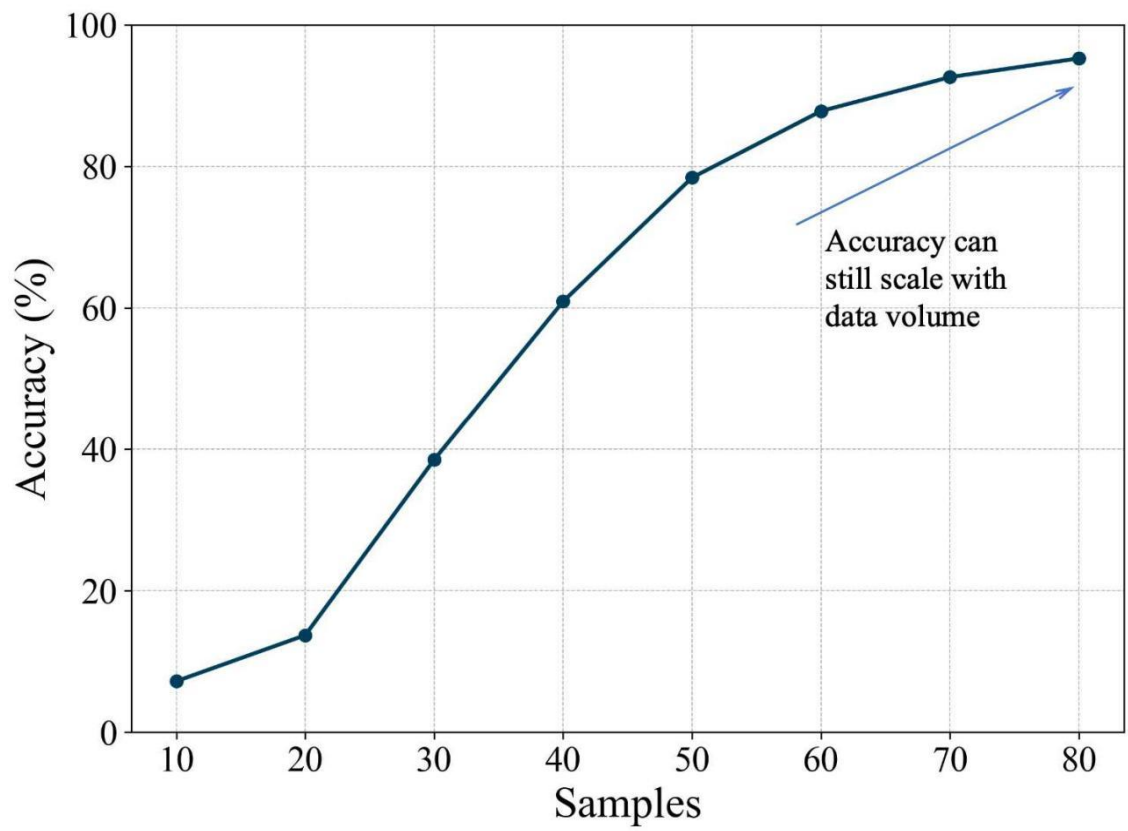

**Supplementary Table 1 | Performance comparison with state-of-the-art silent speech systems.**

| Reference | Accuracy | FLOPS Per Inference | Number of Channels | Number of Words | Sensor Type |
|---|---|---|---|---|---|
| [1] | 89% | 0.03G | 3 | 5 | EMG Sensor |
| [2] | 85.24% | 0.03G | 4 | 5 | EMG Sensor |
| [3] | 87.53% | 0.25G | 8 | 100 | Strain Sensor |
| [4] | 76.7% | 0.1G | 1 | 6 | Strain Sensor |
| [5] | 54.6% | 0.3G | 1 | 15 | Strain Sensor |
| [6] | 99% | 1G | 1 | 5 | Strain Sensor |
| **This Work** | 95.25% | 0.09G | 1 | 20 | Strain Sensor |

**Supplementary Table 2 | Detailed information of the participants.**

| Participant | Gender | Voice-Related Diseases | Age | Average Speaking Decibel Level | Experiment Participation |
|---|---|---|---|---|---|
| Participant 1 | Male | None | 23 | 54 | Main Experiment |
| Participant 2 | Male | None | 24 | 47 | Main Experiment |
| Participant 3 | Female | None | 27 | 42 | Main Experiment |
| Participant 4 | Female | None | 23 | 46 | Generalization Experiment |

**Supplementary Table 3 | Vocabulary details of the datasets.**

| Dataset 1 (20 frequently used words) | | Dataset 2 (10 confusing words) | Dataset 3 (5 long words with different reading speed) |
|---|---|---|---|
| Time | Be | Book | Cambridge |
| Year | Have | Look | University |
| People | Do | Dessert | Electrical |
| Way | Will | Desert | Engineering |
| Man | Say | Record (Noun) | Division |
| Day | Would | Record (Verb) | |
| Thing | Can | Sheep | |
| Child | Get | Ship | |
| Mr | Make | Metal | |
| Government | Go | Medal | |

**Supplementary Table 4 | Hyperparameters for the baseline model.**

|  | Optimal hyperparameters of the baseline model |
|---|---|
| Optimizer | Adam |
| Learning rate | 0.0003 |
| Weight decay | 0.00001 |
| Epochs | 200 |
| Batch size | 32 |